\documentclass[final,twocolumn,prl,aps,floats,amsfonts,amssymb,showpacs]{revtex4}

\usepackage{amsmath}
\usepackage{graphicx}
\begin{document}

\title{Subdiffusion and cage effect in a sheared granular material}
\author{G. Marty and O. Dauchot}

\affiliation{SPEC/DRECAM/DSM/CEA Saclay and CNRS, URA2464, F-91190 Gif sur Yvette Cedex, France}
\date{Received 29 June 2004}

\begin{abstract}
We investigate experimentally the diffusion properties of a bidimensional bidisperse dry granular material under quasistatic cyclic shear.The comparison of these properties with results obtained both in computer simulations of hard spheres systems and Lenard-Jones liquids and experiments on colloidal systems near the glass transition demonstrates a strong analogy between the behaviour of granular matter and  these systems. More specifically, we study in detail the cage dynamics responsible for the subdiffusion in the slow relaxation regime, and obtain the values of relevant time and length scales.
\end{abstract}


\maketitle

\vspace{0.1cm}

Glass forming systems exhibit many intriguing properties, and their study has generated a large amount of theoretical as well as experimental work. One of the main features of their dynamics is what is usually called the cage effect, which accounts for the different relaxation mechanisms \cite{ediger,angell}: at short times, any given particle is trapped in a confined area by its neighbours, which form the so called effective cage, leading to a slow dynamics; at sufficiently long times, the particle has managed to leave its cage, so that it is able to diffuse through the sample by successive cage changes, resulting in a faster relaxation. These mechanisms define the $\beta$ and $\alpha$ regimes.

Many experiments and simulations have been performed to study this scenario. In particular, the understanding of the nature of cages require microscopic informations which have been essentially obtained in computer simulations of hard spheres systems and Lennard-Jones liquids (e.g. \cite{doliwaPRL,doliwaJPhys,allegrini,hurley,donati}). A suitable way to extract these informations in laboratory experiments consists in using systems undergoing a glass transition composed of sufficiently large particles so that it is possible to follow them through direct observation. The main example of this method is the breakthrough study of colloidal particles near the glass transition by confocal microscopy, realized by Weeks {\it{et al.}} \cite{weeksScience,weeksPRL,weeksChemPhys}, who first observed experimentally the cage effect in real space.

Beside,  especially since the crucial experiments of the Chicago group \cite{jaeger,nowak}, it is widely supposed that dense granular matter could be considered as an analog of glassy systems, albeit a rather special one, since it is athermal \cite{kurchan}. Granular systems also undergo a jamming transition which shares many properties with the glass transition, arising the possibility of a unified description \cite{nagel,liu,ohern}. Then, granular materials could represent a simple way to perform accurate measurements and understand the nature of cages (since grains can be relatively large and then quite easy to follow through direct imaging, \cite{pouliquen}) provided one checks that the analogy noticed at the scale of the sample is confirmed by a precise study of the diffusion properties at the grain scale.

In this paper, we show that a very simple system such as a bidimensional bidisperse dry granular material submitted to a quasistatic cyclic shear indeed behaves the same way as glassy systems in the sense that its diffusion properties evaluated with the {\it{same}} tools as in numerical studies of glasses \cite{doliwaPRL,doliwaJPhys,hurley} behave identically (except for the initial thermal regime which obviously does not exist in a granular material). As a result, we also characterize the cage effect and the dynamical heterogeneities.\\

\begin{figure}[h]
\begin{center}
\includegraphics[width=8.5cm,clip]{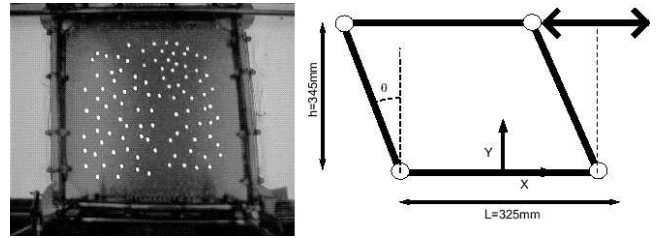}
\caption{\label{disp} Experimental setup (left: picture; right: scheme).\vspace{-0.4cm}}
\end{center}
\end{figure}

The experimental setup is as follows (fig \ref{disp}) : a bidimensional, bidisperse granular material, composed of about 6.000 metallic cylinders of diameter 4 and 5mm in equal proportions, is sheared quasistatically ($I=\dot{\gamma}d/\sqrt{P/\rho}=10^{-5}$; see \cite{gdr}) in an horizontal deformable parallelogram. The shear is periodic, with a shear rate $\dot{\gamma}=1.5^{\circ}.s^{-1}$ and an amplitude $\theta_{max}=\pm 10^{\circ}$. The volume accessible to the grains is maintained constant by imposing the height of the parallelogram, so that the volume fraction is a constant ($\Phi=0.86$). We follow a sample of 500 of the grains with a CCD camera which takes a picture of the material each time the system is back to its initial position ($\theta=0$). The unit of time is then one cycle, a whole experiment lasting 10.000 cycles. The unit of length is chosen to be the mean particle diameter $d$.

\begin{figure}[h]
\begin{center}
\includegraphics[width=8.5cm,clip]{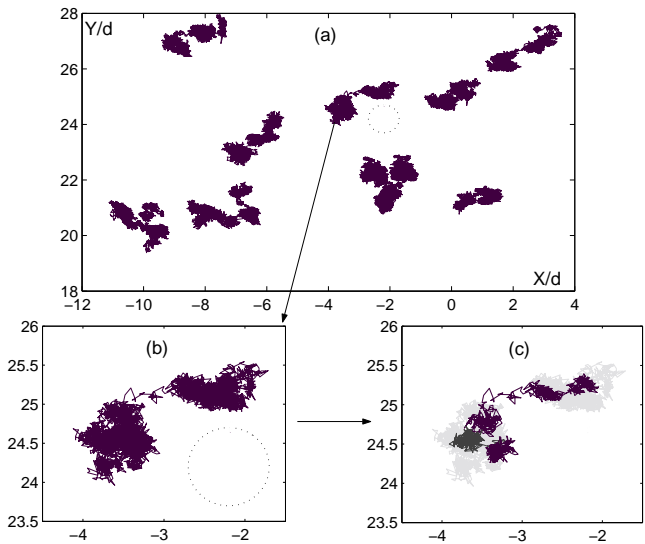}
\caption{\label{traj} (a)some tracers trajectories (b)a typical trajectory; the circle indicates the particle size (c)grey: the same trajectory; black:part of the trajectory, showing the existence of cages at a scale of order 0.3 diameter.}
\end{center}
\end{figure}

Fig \ref{traj}b shows a typical trajectory: the particle spends most of its time confined in a well defined area and sometimes escapes during rare and brief events. In the following, we will refer to this behavior as "cage effect".

\begin{figure}[h]
\begin{center}
\includegraphics[width=8.5cm,clip]{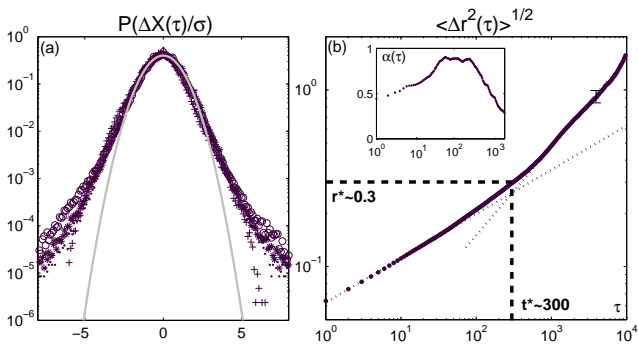}
\caption{\label{stat} (a) pdf of $\Delta X(\tau)/\sigma$ for $\tau=1(.),10(\ast),100(\circ),1000(+)$; the solid line is the gaussian distribution (b) $\sigma(\tau)=\sqrt{\langle \Delta r^2(\tau) \rangle}$; dotted lines show the slopes $1/4$ and $1/2$; dashed lines indicate the position of the crossover which determines $r^{\ast}$ and $t^{\ast}$ (inset: non-gaussian parameter $\alpha(\tau)$). }
\end{center}
\end{figure}

In order to precise the nature of this motion, we first study the statistical properties of the displacements during $\tau$ time steps : $\Delta X(\tau)=X(t+\tau)-X(t)$. The pdf of these displacements are presented on fig \ref{stat}a, for $\tau=$1, 10, 100, 1000. They are caracteristic of interminent dynamics, with fat tails compared with the gaussian case, and then can be interpreted as the signature of the cage effect, which sidesteps the relative proportion of small and large events.

The root mean square displacement $\sigma(\tau)=\sqrt{\Delta r^2(\tau)}$ presents two regimes (fig \ref{stat}b) : at short times, the dynamics is subdiffusive (logarithmic slope $1/4$), which can be interpreted as the result of the trapping of the particles in cages during the $\beta$-relaxation, while it becomes diffusive (logarithmic slope $1/2$) at long times, when particles have succeeded in escaping from the cages, leading to the $\alpha$-relaxation. The crossover between the two regimes can then be considered as a measure of the cage size $r^{\ast}$ and cage lifetime $t^{\ast}$ (see fig \ref{stat}b). Here, we find $r^{\ast}\simeq 0.3$ and $t^{\ast}\simeq 300$. Note that this value of $t^{\ast}$ lies at the end of the range where the non-gaussian parameter $\alpha$, defined as $\alpha(\tau)=\langle \Delta X^4 \rangle / (3\langle \Delta X^2 \rangle ^2)-1$, is maximum (inset of fig \ref{stat}b). This can be understood this way : on timescales much shorter than $t^{\ast}$, as time grows, the particles explore their cage, being statistically more and more influenced by cage borders; when time approaches $t^{\ast}$, they have explored the whole cage and have constructed their statistics, so that $\alpha$ does not change significantly; on timescales longer than $t^{\ast}$, they diffuse from cage to cage, and are then less and less influenced by the effect of trapping, so that $\alpha$ decreases.

Note that the value of $r^{\ast}$ is smaller than the one which could be infered by direct reading of particles trajectories (fig \ref{traj}b). To make this clearer, we then choose to plot only a part of the trajectory. The result is shown on fig \ref{traj}c. It appears that a region that we would visually describe as one cage is in fact the superposition of subcages, each of them having a size of about 0.3, i.e. $r^{\ast}$.

\begin{figure}[h]
\begin{center}
\includegraphics[width=8.5cm,clip]{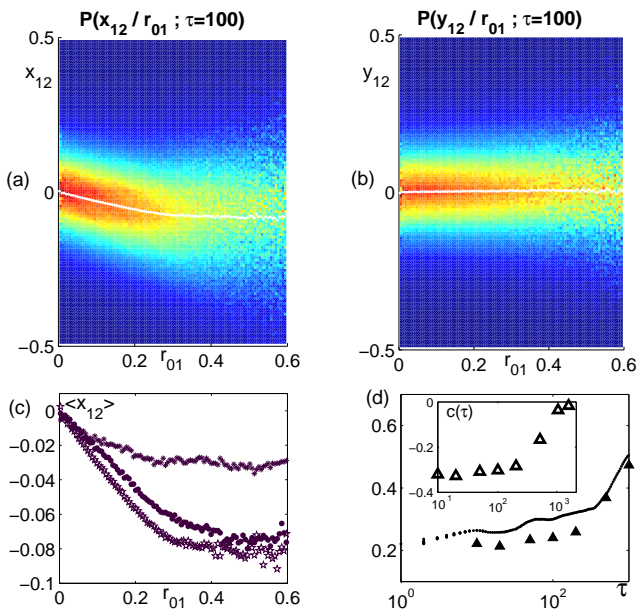}
\caption{\label{pentes} (a) and (b) Conditionnal probabilities (in colorscale) $P(x_{12}\vert r_{01} ;\tau)$ and $P(y_{12}\vert r_{01} ;\tau)$ (see text); the white traces are the mean values $\langle x_{12} \rangle$ and $\langle y_{12} \rangle$ (c) $\langle x_{12} \rangle$ for different values of $\tau$ ( from bottom to top: $\tau=100,300,500$) (d) plot of $\delta(\tau)(.)$ and $\delta_{est}(\tau)(\Delta)$ (inset: slope $c(\tau)$).\vspace{-0.5cm}}
\end{center}
\end{figure}

Following Doliwa \& Heuer \cite{doliwaPRL,doliwaJPhys}, we now turn our attention to the conditional probability $P(x_{12}\vert r_{01} ;\tau)$ (resp $P(y_{12}\vert r_{01} ;\tau)$), which represents the probability distribution of the projection $x_{12}$ (resp. $y_{12}$) of the motion during a time interval $\tau$ along (resp. orthogonaly to) the direction of the motion during the previous time interval, under the condition that the length of the motion during the previous interval has the value $r_{01}$. Results are shown in fig \ref{pentes}. This quantity contains many informations \cite{doliwaPRL}: (i) at a given $r_{01}$, the distributions are symmetric around their mean value ; (ii) the distributions are not gaussian ; (iii) the mean value of $y_{12}$ is 0, while the mean value of $x_{12}$ is always negative ; (iv) if we focus on the evolution of $\langle x_{12} \rangle$ with $r_{01}$, we observe two regions : for $r_{01}<0.3$, there is a linear relation between $\langle x_{12} \rangle$ and $r_{01}$ ($\langle x_{12} \rangle = c(\tau)r_{01}$, $c(\tau)<0$), wherehas for $r_{01}>0.3$, $\langle x_{12} \rangle$ is a constant. Note that the value of $r^{\ast}$ measured using property (iv), i.e. by the localisation of the crossover between the linear and constant evolutions of $\langle x_{12} \rangle$ with $r_{01}$, is the same as the one extracted from the measurement of $\sigma(\tau)$.

The slope $|c(\tau)|$ decreases with $\tau$(fig \ref{pentes}c,d), and is approximately related to the logarithmic slope $\delta(\tau)$ of $\sigma(\tau)$ by  \cite{doliwaPRL,doliwaJPhys}:
\begin{center}
$\delta_{est}(\tau)=0.5+\ln\lbrack 1+c(\tau) \rbrack /\ln4$
\end{center}
Fig \ref{pentes}d shows both $\delta(\tau)$ computed directly from $\sigma(\tau)$ and the one computed using this formula. One sees that the behavior is well reproduced, despite a little offset which might be due to the approximations made in the calculation of $\delta_{est}$.

\begin{figure}[h]
\begin{center}
\includegraphics[width=8.5cm,clip]{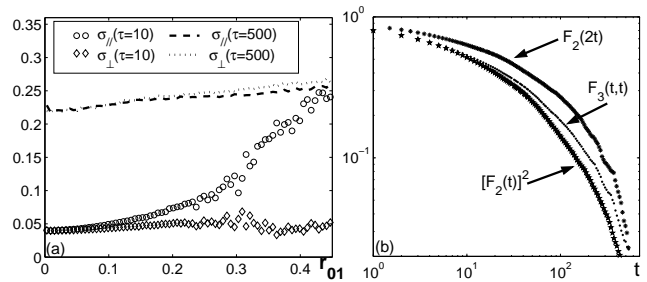}
\caption{\label{sigmastructure} (a) Widths of the distribution of $x_{12}$ ($\sigma_{//}$) and $y_{12}$ ($\sigma_{\perp}$) versus $r_{01}$ for $\tau=10$ and $\tau=500$ (b) $[F_2(t)]^2$, $F_2(2t)$ and $F_3(t,t)$ (see text); at short times, $F_3(t,t)= [F_2(t)]^2$, and at long times, $F_3(t,t)= F_2(2t)$.\vspace{-0.5cm}}
\end{center}
\end{figure}

To go further in the interpretation of these distributions, we extract their widths $\sigma_{//}$ and $\sigma_{\perp}$. Their evolution with $r_{01}$ is shown on fig \ref{sigmastructure}a, for two different values of the time interval $\tau$. First, we note the increase of $\sigma_{//}$ with $r_{01}$. This means that large steps are more likely for particles which  moved farther during the previous interval. It is not the case in a purely diffusive process, since large events are statistical effects, with an occurance which is not related to the length of the previous step, making all the particles to be equivalent. Here, since particles which move farther are the ones which were already making large steps, this shows the existence of a population of fast particles, which is a typical feature of glass forming systems, as pointed out for example in \cite{doliwaPRL,hurley,weeksScience}. Second, we see that for short time intervals $\tau$, the increase of $\sigma_{//}$ is larger than the one of $\sigma_{\perp}$. This reflects the anisotropy of the motion, like stringlike cooperation observed numericaly by Donati {\it{et al.}} \cite{donati}. Both effects concern movements on short timescales, since they tend to disappear as we increase the time interval $\tau$ (fig \ref{sigmastructure}a).

At this point, we can give a partial conclusion about the diffusion properties of the system. The fact that the linear regime of $\langle x_{12}\rangle (r_{01})$ ends at the same value $r_{01}\simeq 0.3$ for any values of $\tau<t^{\ast}$ indicates that, in this regime, it is necessary to describe the system as driven by spatial constraints. For displacements smaller than $r^{\ast}$, the larger a step the more anticorrelated is the following step, as shown by the negative values of $c(\tau)$, which reflects the systematic backdragging effect experienced by the particle in its cage. For displacements longer than $r^{\ast}$, a cage rearrangement has occured, so that the anticorrelation does not increase any more. The constancy of $\langle x_{12}\rangle$ at this saturation value shows a memory of the fact that part of the trajectory was made in a cage.$|c(\tau)|$ decreases with $\tau$, i.e. cage effect becomes weaker, which shows that cages relax and adapt to the new position of the enclosed particles. The fact that the typical relaxation time is $t^{\ast}$ is justified by the need for the particles forming the cage to escape their own ones in order to adapt \cite{doliwaPRL}. On timescales longer than $t^{\ast}$, these effects disappear, which is the result of the increasing number of particles which have undergone a rearrangement.

We now discuss the dynamical heterogeneities by considering multitime correlation functions. Let us define the three quantities $F_2(t)=\langle cos(\vec{q}.\vec{r}_{01})\rangle$, $F_2(2t)=\langle cos(\vec{q}.\vec{r}_{02})\rangle$ and $F_3(t,t)=\langle cos(\vec{q}.\vec{r}_{12}) cos(\vec{q}.\vec{r}_{01}) \rangle$, where the vector $\vec{r}_{ij}$ is the displacement observed between the times $i\times t$ and $j\times t$ : $\vec{r}_{ij}=\vec{r}(j\times t)-\vec{r}(i\times t)$. It has been shown that one can decide wether the dynamic is heterogeneous or homogeneous by comparing $F_3(t,t)$ with $F_2(2t)$ and $[F_2(t)]^2$ respectively \cite{heuerJChemPhys,heuerPRE}. This can be understood by considering the definitions of homogeneous and heterogeneous dynamics \cite{heuerJChemPhys}: in the purely homogeneous case, the movements during two subsequent time intervals along a given direction $\vec{q}$ are {\it{not}} correlated in length, whereas in the purely heterogeneous case, they are {\it{only}} correlated in length. Then, in the homogeneous case, since a cosine is sign independant, both terms of the product are uncorrelated, so that one can factorize $F_3(t,t)$ and obtain $[F_2(t)]^2$. In the heterogeneous case, one can replace $cos(\vec{q}.\vec{r}_{12}) cos(\vec{q}.\vec{r}_{01})$ by $cos(\vec{q}.\vec{r}_{12}+\vec{q}.\vec{r}_{01})+sin(\vec{q}.\vec{r}_{12})sin(\vec{q}.\vec{r}_{01})$. As in this case the signs are not correlated, the mean of the second term must be $0$, so that we are left with $F_2(2t)$. Note that for a random walk, both equalities are fulfilled, so that $F_2(2t)=[F_2(t)]^2$, i.e. the relaxation is exponential. These functions are presented in fig \ref{sigmastructure}b, where we have chosen $q=2\pi/r^{\ast}$. One sees that at short times ($t\leq 10$) the dynamics is mainly homogeneous ($F_3(t,t)= [F_2(t)]^2$), and then slowly evolves toward an heterogeneous dynamics as time grows.

\begin{figure}[h]
\begin{center}
\includegraphics[width=8.5cm,clip]{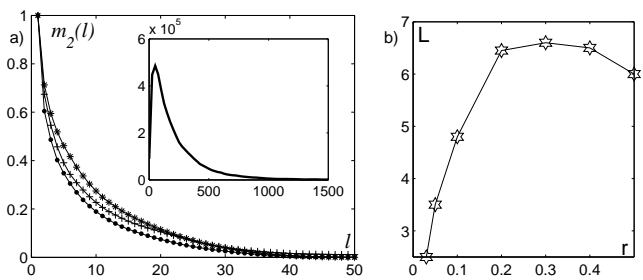}
\caption{\label{m2} (a) second moment $m_2(l)$ of the relaxation time distribution, for different values of the cutoff distance $r$ ($0.1(.)$, $0.3(\ast)$, $0.5(+)$); the dependence of these curves on $r$ is not monotonic (inset: relaxation time distribution for $r=0.3$) (b) characteristic length $L$; it has a maximum $L^{\ast}$ for $r\simeq r^{\ast}$.\vspace{-0.5cm}}
\end{center}
\end{figure}

To better characterize the cooperation in the system, we use a convenient tool proposed by Hurley {\it{et al.}} \cite{hurley}, based on relaxation times. For a particle $i$, the relaxation time $T_i(r)$ is defined as the time needed by the particle to reach a given distance $r$ for the first time. The distribution of these relaxation times is shown in the inset of fig \ref{m2}a, for $r=0.3$. The idea of Hurley {\it{et al.}} is the following: if one defines $T_{i,l}(r)$ as the mean relaxation time of the particles contained in a circle of radius $l$ centered on particle $i$, then the study of the difference between $T_{i,l}(r)$ and $T_{av}$ (where $T_{av}$ is the mean relaxation time calculated over {\it{all}} the particles) should give some informations about the typical length $L$ over which cooperative effects take place. The simplest quantity to compute is then the second moment (in the notations of \cite{hurley}):
\begin{displaymath}
m_2(l)=\frac{\langle (T_{i,l}-T_{av})^2 \rangle}{\langle (T_{i,1}-T_{av})^2 \rangle}
\end{displaymath}

$m_2(l)$ is plotted on fig \ref{m2}, for different values of $r$. One clearly sees that the typical length on which $m_2$ decreases has a maximum around $r\simeq 0.3$. To quantify this, we plot $L$ (defined as the integral over $l$ of $m_2$) versus $r$ and obtain the curve of fig \ref{m2}b. Two important informations can be deduced: first, as we already noticed on the curves of $m_2(l)$, $L$ as a maximum for $r\simeq 0.3$, i.e. the typical cage size. This means that cage rearrangements are phenomena which imply more cooperativity than the dynamics at other scales (a complete discussion is given in ref \cite{hurley}). Second, we obtain a value for this typical length $L$ which is, at its maximum, $L^{\ast}\simeq 7.5$ particle diameters. We then see that cage rearrangements are highly cooperative phenomena. This, added to the small value of $r^{\ast}$, shows that the picture of a particle escaping from its nearest neighbours is definitely not adapted. Instead, these events are subtle and complicated rearrangements, involving a large number of particles. One remaining open question is the nature of the slow phenomenon to which the cage dynamics participate, since, contrary to the experiment of Pouliquen {\it{et al.}} \cite{pouliquen} on granular compaction, the present one is not submitted to gravity and has a constant volume. Such a question as already been raised by Kabla \& Debr\'egeas \cite{kabla} in the case of a {\it{gently}} vibrated pile.

The first conclusion of this experimental work is that scenari imagined to describe glass forming systems apply to the peculiar case of dense granular materials, giving a precise sense to the strong analogy between these two fields and conforting the application of theoretical ideas inherited from statisical physics of glassy systems to granular matter. The second important result is that the peculiar dynamics observed in glassy materials still exists in a system to which an agitation far from a thermal noise is provided. Also, this experiment shows that a simple granular system as the one described in this paper may be an efficient laboratory model to look for  the microscopic phenomena involved in cage rearrangements, which are still poorly understood, but of capital interest in the understanding of jamming and glass transitions.

We wish to thank M. Wyart, J.P. Bouchaud and E. Bertin for helpfull discussions and C. Gasquet and V. Padilla for very efficient technical support.

\end{document}